# Private Quantum Coding for Quantum Relay Networks

Laszlo Gyongyosi[*], Sandor Imre

*Quantum Technologies Laboratory*
*Department of Telecommunications*
*Budapest University of Technology and Economics*
2 Magyar tudosok krt, H-1111 Budapest, Hungary
[*]*gyongyosi@hit.bme.hu*

*Abstract*— The relay encoder is an unreliable probabilistic device which is aimed at helping the communication between the sender and the receiver. In this work we show that in the quantum setting the probabilistic behavior can be completely eliminated. We also show how to combine quantum polar encoding with superactivation-assistance in order to achieve reliable and capacity-achieving private communication over noisy quantum relay channels.

## I. Introduction

The superactivation effect originally was intended to use zero-capacity quantum channels for communication. In this paper we firstly demonstrate that the superactivation effect can be exploited for a completely different propose. We will use the superactivation-effect to construct deterministic quantum relay encoders.

The *superactivation* is an extreme violation of the additivity of quantum channel capacities and enables the use of zero-capacity quantum channels for communication [5], [12], [14], [16-19]. Here we show, that this effect can also be exploited in the construction of efficient quantum relay encoders using quantum polar codes to achieve the private classical capacity of the quantum relay channel. The channel polarization scheme introduced by Arikan [3] makes it possible to achieve the symmetric capacity of a noisy communication channel. The polar coding technique was extended to secure communication over classical channels by Mahdavifar and Vardy [4]. The quantum polar coding scheme was studied by Wilde and Guha [6] and by Renes et al. [7]. Later, the results of [7] were extended by Wilde and Renes [13], [21] for arbitrary quantum channels.

The relay encoding scheme for classical communication channels was introduced by Cover and Gamal [1]. The relay encoding with classical polar coding was studied by Andersson et al. [2], using their nested polar relay channel codes. Here we show that relay encoding can also be used in the quantum setting to achieve enhanced private communication between the sender and receiver. Our two main goals can be summarized as follows: First, using quantum polar coding we would like to maximize the transmittable private classical information over the quantum relay channel. Second, we would like to prove that using superactivation-assistance the reliability of the quantum relay encoder can be maximized and the probabilistic behavior can be completely eliminated. The relay encoder is placed between the Alice and Bob and intended to help Bob receiving messages from Alice [1]. The channel which contains an $\mathcal{E}_2$ relay encoder between Alice and Bob is called the *relay channel* [1] (see Fig. 1.). In case of a *degraded* relay channel, the relay receiver $\mathcal{E}_2$ works better than Bob's receiver $\mathcal{D}$ and the relay encoder can cooperate with original encoder $\mathcal{E}_1$ to help to decode the message on Bob's side [1], [2].

### A. The Quantum Channel

The map of the quantum channel can be expressed with a special representation called the *Kraus Representation* [15], [20]. For a given input system $\rho_A$ and the quantum channel $\mathcal{N}$, this representation can be expressed as

$$\mathcal{N}(\rho_A) = \sum_i N_i \rho_A N_i^\dagger, \qquad (1)$$

where $N_i$ are the Kraus operators, and $\sum_i N_i^\dagger N_i = I$ [15]. The isometric extension [15], [20] of $\mathcal{N}$ by means of the *Kraus Representation* can be expressed as

$$\begin{aligned}\mathcal{N}(\rho_A) &= \sum_i N_i \rho_A N_i^\dagger \to U_{A \to BE}(\rho_A) \\ &= \sum_i N_i \otimes |i\rangle_E.\end{aligned} \qquad (2)$$

The action of the quantum channel $\mathcal{N}$ on an operator $|k\rangle\langle l|$, where $\{|k\rangle\}$ is an orthonormal basis, also can be given in operator form using the Kraus operator $N_{kl} = \mathcal{N}(|k\rangle\langle l|)$. By exploiting the property $UU^\dagger = P_{BE}$, for the input quantum system $\rho_A$

$$\begin{aligned} U_{A \to BE}(\rho_A) &= U\rho_A U^\dagger \\ &= \left(\sum_i N_i \otimes |i\rangle_E\right)\rho_A\left(\sum_j N_j^\dagger \otimes \langle j|_E\right) \\ &= \sum_{i,j} N_i \rho_A N_j^\dagger \otimes |i\rangle\langle j|_E. \end{aligned} \quad (3)$$

If we trace out the environment, we get

$$Tr_E\left(U_{A \to BE}(\rho_A)\right) = \sum_i N_i \rho_A N_i^\dagger. \quad (4)$$

The *isometric extension* of the quantum channel $\mathcal{N}$ is simply the unitary representation of the channel

$$\mathcal{N}: U_{A \to BE}, \quad (5)$$

where $A$ is the input system, $B$ is the channel output of the unitary transformation, and $E$ is the environment of the quantum channel. By applying $U_{A \to BE}$ to the input quantum system $\rho$, we have

$$Tr_E\left(U_{A \to BE}(\rho_A)\right) = \mathcal{N}(\rho_A). \quad (6)$$

The isometry of a unitary map $U$ holds two attributes. First, $U^\dagger U = I$, that is, it behaves just like an ordinary unitary transformation, and $UU^\dagger = P_{BE}$, where $P_{BE}$ is a projector applied to the joint output $BE$ of the quantum channel [6], [10], [15].

### B. The Symmetric Capacity of Quantum Channels

An important property of the *symmetric* capacity, that there is *no maximization* in the mutual information function, since the distribution of the input states is assumed to be *uniform* [3], [6], [13], [21]. Assuming the uniformly distributed $A$ classical input system with $a \in \{0,1\}$, the channel output system $\boldsymbol{B}$ (described by quantum states $\sigma_0^B, \sigma_1^B$) with respect to $A$ is defined by [6]

$$\sigma^{AB} = \frac{1}{2}|0\rangle\langle 0|^A \otimes \sigma_0^B + \frac{1}{2}|1\rangle\langle 1|^A \otimes \sigma_1^B. \quad (7)$$

For a quantum channel $\mathcal{N}$ with input system $A$ and output system $B$, the *symmetric* classical capacity is equal to the symmetric quantum mutual information $I(A:B)$ [6], [13], [21],

$$C_{sym}(\mathcal{N}) = I(A:B). \quad (8)$$

(*Note*: the $I(A:B)$ symmetric quantum mutual information is additive for a quantum channel.) The result in (8) further can be evaluated as [6]

$$C_{sym}(\mathcal{N}) = \mathrm{S}\left((\sigma_0^B + \sigma_1^B)/2\right) - \mathrm{S}(\sigma_0^B)/2 - \mathrm{S}(\sigma_1^B)/2. \quad (9)$$

For the *symmetric* private classical capacity [11], [21] the same condition holds, i.e., there is *no maximization* needed because the inputs are uniformly distributed and the channels between Alice and Bob, and Alice and Eve are symmetric:

$$\begin{aligned} P_{sym}(\mathcal{N}) &= \lim_{n \to \infty} \frac{1}{n} P_{sym}^{(1)}(\mathcal{N}^{\otimes n}) \\ &= \lim_{n \to \infty} \frac{1}{n}(I(A:B) - I(A:E)), \end{aligned} \quad (10)$$

where $P_{sym}^{(1)}(\mathcal{N})$ is the *single-use symmetric* private classical capacity (*symmetric private information*)

$$P_{sym}^{(1)}(\mathcal{N}) = I(A:B) - I(A:E), \quad (11)$$

and $I(A:E)$ is the symmetric quantum mutual information function between Alice and Eve.

## II.    THE POLAR ENCODING SCHEME

In our scheme we use polar encoding to achieve the private classical capacity. Our encoding scheme is summarized as follows. Alice encodes her private message $M$ into the phase information using the $X$ basis and then into the amplitude using the $Z$ basis. The phase carries the *data*, while the amplitude is the *key* for the encryption i.e., in our scheme Alice first encodes the phase (data) and then the amplitude (key). Bob applies it in the reverse direction using his successive and coherent decoder, and finally gets $M'$ as follows [7], [21]: he first decodes the *amplitude* (key) information in the $Z$ basis [8-9]. Then Bob continues the decoding with the *phase* information, in the $X$ basis [7], [21]. An important ingredient of our scheme: the successful decoding of the amplitude information (key) is a *necessary but not sufficient* condition for the positive private classical capacity. According to our encoding scheme, the *symmetric private classical capacity* $P_{sym}$ is defined as

$$P_{sym}(\mathcal{N}) = \lim_{n\to\infty} \frac{1}{n}\left(I_{sym.}^{phase}(A:B) - I(A:E)\right)$$
$$= \lim_{n\to\infty} \frac{1}{n}\begin{pmatrix} S\left((\sigma_0^{phase}+\sigma_1^{phase})/2\right) - S\left(\sigma_0^{phase}\right)/2 \\ -S\left(\sigma_1^{phase}\right)/2 - I(A:E) \end{pmatrix},$$

where $S(\cdot)$ is the von Neumann entropy, while $I(A:E) = S(A) + S(E) - S(AE)$ stands for the mutual information function between Alice and Eve, and $I_{sym.}^{phase}(A:B)$ is the symmetric mutual information that can be achieved by the phase information between Alice and Bob [6], [13], [21]. To construct the input polar codeword sets, we use the notation of 'good' $\mathcal{G}(\mathcal{N},\beta)$ and 'bad' $\mathcal{B}(\mathcal{N},\beta)$, where $\beta < 0.5$ [3], [6], [13], [21], channels for the transmission of phase and amplitude (see *Appendix*). The $S_{in}$ set of polar codewords which can transmit private information (both amplitude and phase) is denoted by [4], [7], [13], [21]:

$$S_{in} = \mathcal{G}(\mathcal{N}_{amp},\beta) \cap \mathcal{G}(\mathcal{N}_{phase},\beta). \quad (12)$$

All of other input codewords cannot transmit private classical information. These codewords are defined by the set $S_{bad}$ as follows [7], [13], [21]:

$$S_{bad} = \left(\mathcal{G}(\mathcal{N}_{amp},\beta) \cap \mathcal{B}(\mathcal{N}_{phase},\beta)\right) \cup$$
$$\left(\mathcal{B}(\mathcal{N}_{amp},\beta) \cap \mathcal{G}(\mathcal{N}_{phase},\beta)\right) \cup \quad (13)$$
$$\left(\mathcal{B}(\mathcal{N}_{amp},\beta) \cap \mathcal{B}(\mathcal{N}_{phase},\beta)\right).$$

From set $S_{bad}$, we define the completely useless codewords as

$$\mathcal{B} = \mathcal{B}(\mathcal{N}_{amp},\beta) \cap \mathcal{B}(\mathcal{N}_{phase},\beta), \quad (14)$$

while the 'partly good' (i.e., can be used for non private classical communication) input codewords will be denoted by

$$\mathcal{P}_1 = \mathcal{G}(\mathcal{N}_{amp},\beta) \cap \mathcal{B}(\mathcal{N}_{phase},\beta) \quad (15)$$

and

$$\mathcal{P}_2 = \mathcal{B}(\mathcal{N}_{amp},\beta) \cap \mathcal{G}(\mathcal{N}_{phase},\beta). \quad (16)$$

The working process of the polar-encoding based quantum relay encoder $\mathcal{E}_2$ can be summarized as follows: The first encoder $\mathcal{E}_1$ encodes the *phase* information into the codeword $A$ and then sends it to $\mathcal{E}_2$, using set $\mathcal{G}(\mathcal{N}_{phase},\beta)$. In the next step, the quantum relay encoder $\mathcal{E}_2$ with probability $p_{\mathcal{E}_2}$ adds the *amplitude* information to the phase information, using polar codes from the set $\mathcal{G}(\mathcal{N}_{phase},\beta) \setminus \mathcal{P}_2$, and then sends it to Bob. Otherwise, with probability $(1 - p_{\mathcal{E}_2})$ it leaves unchanged $\mathcal{G}(\mathcal{N}_{phase},\beta)$ and transmits to Bob. The problem with the quantum relay encoder $\mathcal{E}_2$ is that it is unreliable since it works in a probabilistic way, which also makes the channel $\mathcal{N}_{\mathcal{E}_2\mathcal{D}}$ too noisy. The input codewords that can be used with our encoding scheme are already defined in (12) by the set of $S_{in}$. However, (13) can be split into more sets, and several security issues can be derived from these polar codewords [4], [7], [13], [21]. The quantum channel $\mathcal{N}$ is called *degraded* if the channel $\mathcal{N}_{Eve}$ between Alice and Eve is "noisier" than the channel $\mathcal{N}_{Bob}$ between Alice and Bob. Using (14), for a non-degraded eavesdropper, $P_{sym}$ can be expressed as [4], [13], [21]

$$P_{sym}(\mathcal{N}) = \lim_{n\to\infty} \frac{1}{n}(|S_{in}| - |\mathcal{B}|), \quad (17)$$

which can be extended for the polar codeword sets as follows:

$$P_{sym}(\mathcal{N}) = \lim_{n\to\infty} \frac{1}{n}\begin{pmatrix} |\mathcal{G}(\mathcal{N}_{amp},\beta)| + |\mathcal{G}(\mathcal{N}_{phase},\beta)| - \\ |\mathcal{G}(\mathcal{N}_{amp},\beta) \cup \mathcal{G}(\mathcal{N}_{phase},\beta)| \end{pmatrix}$$
$$- \lim_{n\to\infty} \frac{1}{n}\left[n - |\mathcal{G}(\mathcal{N}_{amp},\beta) \cup \mathcal{G}(\mathcal{N}_{phase},\beta)|\right]$$
$$= \lim_{n\to\infty} \frac{1}{n}\left(|\mathcal{G}(\mathcal{N}_{amp},\beta)| + |\mathcal{G}(\mathcal{N}_{phase},\beta)| - n\right). \quad (18)$$

The input codewords from $\mathcal{P}_1$ and $\mathcal{P}_2$ cannot be used to transmit classical information *privately*, since these codewords do not satisfy our requirements on the encoding of private information (only set $S_{in}$ is allowed in our encoding scheme.).

### III. THEOREMS AND PROOFS ON THE ENCODING SCHEME

The main results on the security of the proposed quantum polar coding scheme are summarized in Theorems 1 and 2.

**Theorem 1.** *Assuming a degraded quantum channel $\mathcal{N}_{Eve}$, the following $P_{sym}$ symmetric private classical capacity can be achieved over the quantum channel $\mathcal{N}_{Bob}$:*

$$P_{sym} = \lim_{n \to \infty} \frac{1}{n}(|S_{in}|) \qquad (19)$$
$$= \lim_{n \to \infty} \frac{1}{n}|\mathcal{G}(\mathcal{N}_{amp},\beta) \cap \mathcal{G}(\mathcal{N}_{phase},\beta)|.$$

*Proof.* According to our results, if Eve's channel is degraded, then for the $P_{sym}$ achievable private capacity (19) will hold for the rate $R_{sym}$. Assuming $\beta < 0.5$, that

$$C(\mathcal{N}_{Eve}) = \lim_{n \to \infty} \frac{1}{n}(|\mathcal{P}_1| + |\mathcal{P}_2|). \qquad (20)$$

Combing this result with (36), we get

$$C(\mathcal{N}_{Eve}) = \lim_{n \to \infty} \frac{1}{n}(|\mathcal{P}_1|) \qquad (21)$$

and

$$C(\mathcal{N}_{Bob}) = 1 - C(\mathcal{N}_{Eve})$$
$$= 1 - \lim_{n \to \infty} \frac{1}{n}(|\mathcal{P}_1|). \qquad (22)$$

The result obtained in (22) can be rewritten as follows [4], [21]:

$$C(\mathcal{N}_{Bob}) = \lim_{n \to \infty} \frac{1}{n}(|S_{in}| \cup |\mathcal{P}_2|) = \lim_{n \to \infty} \frac{1}{n}(|S_{in}|). \qquad (23)$$

According to our polar coding scheme:

$$S_{in} = \left\{ i \in n : B(\mathcal{N}_i) < \frac{1}{n} 2^{-n^\beta} \right\}, \qquad (24)$$

where $\beta < 0.5$, and

$$[n] \setminus S_{in} = \left\{ i \in n : B(\mathcal{N}_i) \geq 1 - \frac{1}{n} 2^{-n^\beta} \right\}. \qquad (25)$$

From the definition of (24) and (25), for the Bhattacharya parameters (analogous to *fidelity* of the channel [4], [6-7], [13], [21], see *Appendix*) of these codewords, we have the following relation. Using the sets as defined in (12) and (14), it follows that

$$\begin{aligned}&\left(\mathcal{B}(\mathcal{N}_{amp},\beta) \cap \mathcal{B}(\mathcal{N}_{phase},\beta)\right) \subseteq \\ &\left(S_{in} \cup \left(\mathcal{B}(\mathcal{N}_{amp},\beta) \cap \mathcal{B}(\mathcal{N}_{phase},\beta)\right)\right) = \varnothing.\end{aligned} \qquad (26)$$

After some steps of reordering, we get that

$$\begin{aligned}&\left(\mathcal{G}(\mathcal{N}_{amp},\beta) \cap \mathcal{B}(\mathcal{N}_{phase},\beta)\right) \\ &\cap \left(\mathcal{B}(\mathcal{N}_{amp},\beta) \cap \mathcal{B}(\mathcal{N}_{phase},\beta)\right) \subseteq \\ &\left(\mathcal{G}(\mathcal{N}_{amp},\beta) \cap \mathcal{B}(\mathcal{N}_{phase},\beta)\right) \\ &\cap \left(S_{in} \cup \left(\mathcal{B}(\mathcal{N}_{amp},\beta) \cap \mathcal{B}(\mathcal{N}_{phase},\beta)\right)\right) = \varnothing.\end{aligned} \qquad (27)$$

From these results follow that,

$$\mathcal{P}_1 \cap (S_{in} \cup \mathcal{P}_2) = \varnothing. \qquad (28)$$

This result also means that the constructed codeword sets $S_{in}$, $\mathcal{P}_1$, $\mathcal{P}_2$, and $\mathcal{B}$ are disjoint sets [4], [13], [21] with relation $|S_{in} \cup \mathcal{P}_1 \cup \mathcal{P}_2| = n$. As follows, if Eve's channel $\mathcal{N}_{Eve}$ is a non-degraded quantum channel [13], then

$$P_{sym}(\mathcal{N}) = \lim_{n \to \infty} \frac{1}{n}|S_{in}| - \lim_{n \to \infty} \frac{1}{n}|\mathcal{B}|, \qquad (29)$$

while, if Eve's channel is degraded quantum channel [13], then

$$\lim_{n \to \infty} \frac{1}{n}|\mathcal{B}| = 0, \qquad (30)$$

which concludes our proof on $P_{sym}(\mathcal{N})$ for a degraded eavesdropper channel [4], [13], [21]:

$$P_{sym}(\mathcal{N}) = \lim_{n \to \infty} \frac{1}{n}|S_{in}|. \qquad (31)$$

∎

Our results on the achievable rate of secret private communication assuming a non-degraded quantum channel $\mathcal{N}_{Eve}$ between Alice and Eve are summarized in Theorem 2.

**Theorem 2.** *Assuming a non-degradable quantum channel $\mathcal{N}_{Eve}$, for the sets $\mathcal{B}(\mathcal{N}_{amp},\beta)$ and $\mathcal{P}_2$, the following relation holds for the rate $R_{sym}$ of the private communication between Alice and Bob:*

$$\begin{aligned}R_{sym} &= \frac{1}{n}(|S_{in}| + |\mathcal{B}| - |\mathcal{B}(\mathcal{N}_{amp},\beta)| + |\mathcal{P}_2|) \\ &= \frac{1}{n}\begin{pmatrix}|\mathcal{G}(\mathcal{N}_{amp},\beta) \cap \mathcal{G}(\mathcal{N}_{phase},\beta)| \\ +|\mathcal{B}(\mathcal{N}_{amp},\beta) \cap \mathcal{B}(\mathcal{N}_{phase},\beta)| \\ -|\mathcal{B}(\mathcal{N}_{amp},\beta)| + |\mathcal{B}(\mathcal{N}_{amp},\beta) \cap \mathcal{G}(\mathcal{N}_{phase},\beta)_2|\end{pmatrix},\end{aligned}$$
*(32)*

*where for polar codeword set $\mathcal{P}_2$*

$$\lim_{n \to \infty} \frac{1}{n}|\mathcal{P}_2| = 0, \qquad (33)$$

*and*

$$\begin{aligned}S_{in} \cup (\mathcal{P}_1 \cup \mathcal{P}_2) &= \mathcal{G}(\mathcal{N}_{amp},\beta) \cup \\ &(\mathcal{B}(\mathcal{N}_{amp},\beta) \cap \mathcal{G}(\mathcal{N}_{phase},\beta)),\end{aligned} \qquad (34)$$

*with $p(M \neq M') = 2^{-n^\beta} + \sum_i B(\mathcal{N}_i)$, where $2^{-n^\beta}$ is the upper bound on set $\mathcal{G}(\mathcal{N}_{amp},\beta)$ and $B$ is the Bhattacharya parameter of the i-th channel $\mathcal{N}_i$.*

*Proof.* Assuming a non-degradable quantum channel $\mathcal{N}_{Eve}$, for the sets $\mathcal{B}(\mathcal{N}_{amp},\beta)$ and $\mathcal{P}_2$, the following relation holds for the rate $R_{sym}$ of the private communication between Alice and Bob [4], [7], [13], [21]:

$$R_{sym} = \frac{1}{n}(|S_{in}| + |\mathcal{B}| - |\mathcal{B}(\mathcal{N}_{amp},\beta)| + |\mathcal{P}_2|)$$
$$= \frac{1}{n}\begin{pmatrix} |\mathcal{G}(\mathcal{N}_{amp},\beta) \cap \mathcal{G}(\mathcal{N}_{phase},\beta)| \\ + |\mathcal{B}(\mathcal{N}_{amp},\beta) \cap \mathcal{B}(\mathcal{N}_{phase},\beta)| \\ - |\mathcal{B}(\mathcal{N}_{amp},\beta)| + |\mathcal{B}(\mathcal{N}_{amp},\beta) \cap \mathcal{G}(\mathcal{N}_{phase},\beta)_2| \end{pmatrix},$$
(35)

where for polar codeword set $\mathcal{P}_2$

$$\lim_{n\to\infty}\frac{1}{n}|\mathcal{P}_2| = 0, \quad (36)$$

with $p(M \neq M') = 2^{-n^\beta} + \sum_i B(\mathcal{N}_i)$, where $2^{-n^\beta}$ is the upper bound on set $\mathcal{G}(\mathcal{N}_{amp},\beta)$ [7], [13], [21] and $B$ is the Bhattacharya parameter of the $i$-th channel $\mathcal{N}_i$.  ∎

For a degradable quantum channel $\mathcal{N}_{Eve}$, $\lim_{n\to\infty}\frac{1}{n}|\mathcal{B}| = 0$ [13], [21], i.e., $R_{sym}$ can be rewritten as follows:

$$R_{sym} = \frac{1}{n}(|S_{in}| - |\mathcal{P}_2| + |\mathcal{P}_2|) = \frac{1}{n}(|S_{in}|)$$
$$= \frac{1}{n}(|\mathcal{G}(\mathcal{N}_{amp},\beta) \cap \mathcal{G}(\mathcal{N}_{phase},\beta)|).$$
(37)

If $\mathcal{N}_{Eve}$ is a non-degraded quantum channel, then for private communication the following sets of polar codewords can be used:

$$|S_{in}| - |\mathcal{B}(\mathcal{N}_{phase},\beta)| = |\mathcal{G}(\mathcal{N}_{amp},\beta) \cap \mathcal{G}(\mathcal{N}_{phase},\beta)| - |\mathcal{B}(\mathcal{N}_{phase},\beta)|. \quad (38)$$

If $\mathcal{N}_{Eve}$ is a degraded quantum channel [13], [21], then

$$|S_{in}| = |\mathcal{G}(\mathcal{N}_{amp},\beta) \cap \mathcal{G}(\mathcal{N}_{phase},\beta)|, \quad (39)$$

i.e., in each case, there exist input quantum polar codeword $s_{in} \in S_{in}$, which satisfies our criteria on the encoding of private information, as stated in (12).

## IV. THE QUANTUM RELAY ENCODER

Our proposed quantum relay encoder $\mathcal{E}_2$ is depicted in Fig. 1. Alice would like to send her *l*-length private message $M$ to Bob. The first encoder $\mathcal{E}_1$ can encode only phase information, while the quantum relay encoder $\mathcal{E}_2$ can encode only amplitude information. The quantum relay encoder $\mathcal{E}_2$ can add the amplitude information to the message $A$ received from $\mathcal{E}_1$ only with success probability $0 < p_{\mathcal{E}_2} < 1$. In the first step, her encoder $\mathcal{E}_1$ outputs the *n*-length *phase* encoded message $A$. The second encoder $\mathcal{E}_2$ gets input on the channel output $B'$, which will be amended with *amplitude* information. The relay quantum encoder $\mathcal{E}_2$ outputs $A'$ to the channel, and Bob will receive message $B$. The goal of the whole structure is to help Bob's encoder $\mathcal{D}$, by the quantum relay encoder $\mathcal{E}_2$ to cooperate with $\mathcal{E}_1$, to send the private message $M$ from Alice to Bob. The quantum relay channel $\mathcal{N}_{\mathcal{E}_1\mathcal{E}_2\mathcal{D}}$, which includes Alice's first encoder $\mathcal{E}_1$ and the relay encoder $\mathcal{E}_2$ is defined as $\mathcal{N}_{\mathcal{E}_1\mathcal{E}_2\mathcal{D}} = \mathcal{N}_{\mathcal{E}_1\mathcal{E}_2}\mathcal{N}_{\mathcal{E}_2\mathcal{D}}$, where $\mathcal{N}_{\mathcal{E}_1\mathcal{E}_2}$ is the quantum channel between encoder $\mathcal{E}_1$ and the quantum relay encoder $\mathcal{E}_2$, while $\mathcal{N}_{\mathcal{E}_2\mathcal{D}}$ is the quantum channel between $\mathcal{E}_2$ and Bob's decoder $\mathcal{D}$.

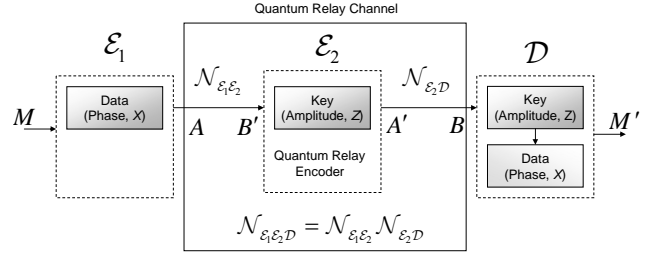

**Fig. 1.** The quantum relay channel with the relay encoder. The first encoder encodes only phase information, the second adds only the amplitude information. The quantum relay encoder is not reliable, it works with success probability $0 < p_{\mathcal{E}_2} < 1$.

For a degradable quantum relay channel $\mathcal{N}_{\mathcal{E}_1\mathcal{E}_2\mathcal{D}}$, the channel $\mathcal{N}_{\mathcal{E}_1\mathcal{D}}$ is noisier than $\mathcal{N}_{\mathcal{E}_1\mathcal{E}_2\mathcal{D}} = \mathcal{N}_{\mathcal{E}_1\mathcal{E}_2}\mathcal{N}_{\mathcal{E}_2\mathcal{D}}$. The $\mathcal{N}_{\mathcal{E}_1\mathcal{E}_2\mathcal{D}}$ quantum relay channel is noisy due to the unreliable quantum relay encoder $\mathcal{E}_2$. The symmetric classical capacity of quantum relay channel $\mathcal{N}_{\mathcal{E}_1\mathcal{E}_2\mathcal{D}}$, assuming encoders $\mathcal{E}_1$ and $\mathcal{E}_2$, and decoder $\mathcal{D}$ can be expressed as

$$\begin{aligned}C_{sym}\left(\mathcal{N}_{\mathcal{E}_1\mathcal{E}_2\mathcal{D}}\right) &= \max_{p(A,A')} \min\left\{I(A,A':B), I(A:B'|A')\right\} \\ &= I(A:B,B'|A') = I(A:B'|A'),\end{aligned}$$
(40)

which is equivalent to the following definition. Let be the classical capacity of the channel between encoder $\mathcal{E}_1$ and quantum relay encoder $\mathcal{E}_2$ is $C_{sym}\left(\mathcal{N}_{\mathcal{E}_1\mathcal{E}_2}\right)$, while between $\mathcal{E}_1$ and $\mathcal{D}$ without quantum relay encoder $\mathcal{E}_2$ is $C_{sym}\left(\mathcal{N}_{\mathcal{E}_1\mathcal{D}}\right)$ and between $\mathcal{E}_2$ and $\mathcal{D}$ is $C_{sym}\left(\mathcal{N}_{\mathcal{E}_2\mathcal{D}}\right)$. Using these channels, the capacity $C_{sym}\left(\mathcal{N}_{\mathcal{E}_1\mathcal{E}_2\mathcal{D}}\right)$ of $\mathcal{N}_{\mathcal{E}_1\mathcal{E}_2\mathcal{D}}$ can be calculated as

$$\begin{aligned}&C_{sym}\left(\mathcal{N}_{\mathcal{E}_1\mathcal{E}_2\mathcal{D}}\right) \\ &= \min\left\{C_{sym}\left(\mathcal{N}_{\mathcal{E}_1\mathcal{E}_2}\right), \left(C_{sym}\left(\mathcal{N}_{\mathcal{E}_1\mathcal{D}}\right) + C_{sym}\left(\mathcal{N}_{\mathcal{E}_2\mathcal{D}}\right)\right)\right\}.\end{aligned}$$
(41)

Our quantum relay encoder differs from the classical relay encoder scheme of [1], [2]. The $\mathcal{E}_2$ quantum relay encoder outputs codeword

$$B = p_{\varepsilon_2}\left(\mathcal{G}(\mathcal{N}_{phase},\beta)\setminus\mathcal{P}_2\right) + \left(1-p_{\varepsilon_2}\right)\mathcal{G}(\mathcal{N}_{phase},\beta),$$

with

$$|B| = p_{\varepsilon_2}\left|\left(\mathcal{G}(\mathcal{N}_{phase},\beta)\setminus\mathcal{P}_2\right)\right|$$
(42)

or

$$|B| = \left(1-p_{\varepsilon_2}\right)\left|\mathcal{G}(\mathcal{N}_{phase},\beta)\right|,$$
(43)

where $p_{\varepsilon_2}$ is the success probability of the quantum relay encoder $\mathcal{E}_2$. Finally, Bob decodes the message using set $\mathcal{G}(\mathcal{N}_{phase},\beta)\setminus\mathcal{P}_2$.

If Bob received $B = \mathcal{G}(\mathcal{N}_{phase},\beta)$ from the quantum relay encoder $\mathcal{E}_2$, then the decoding process fails—this occurs with probability $\left(1-p_{\varepsilon_2}\right)$.

### A. Quantum Relay Encoder with Superactivation-Assistance

As we will prove, using *superactivation-assistance* in the relay channel $\mathcal{N}_{\mathcal{E}_1\mathcal{E}_2\mathcal{D}}$, the reliability of $\mathcal{E}_2$ will be $p_{\varepsilon_2} = 1$; however, the rate of private communication will be lower, which will result in the codeword

$$B^* = \mathcal{G}(\mathcal{N}_{phase},\beta)\setminus\mathcal{P}_2.$$
(44)

with $\left|B^*\right| = \tfrac{1}{2}\left|\left(\mathcal{G}(\mathcal{N}_{phase},\beta)\setminus\mathcal{P}_2\right)\right|$. For the *rate* of private communication, any benefits from the superactivation-assistance can be exploited if and only if $0 < p_{\varepsilon_2} < 0.5$, since in that case $\left|B^*\right| > |B|$. As follows, we will assume an unreliable quantum relay encoder with success probability $0 < p_{\varepsilon_2} < 0.5$. The achievable symmetric classical capacities can be summarized as follows: The symmetric classical capacity between Alice and the quantum relay encoder $\mathcal{E}_2$ is $C_{sym}\left(\mathcal{N}_{\mathcal{E}_1\mathcal{E}_2}\right) = \lim_{n\to\infty}\tfrac{1}{n}\left|\mathcal{G}(\mathcal{N}_{phase},\beta)\right|$. The symmetric classical capacity between Alice and Bob with no relay encoder $\mathcal{E}_2$ assistance is $C_{sym}\left(\mathcal{N}_{\mathcal{E}_1\mathcal{D}}\right) = \lim_{n\to\infty}\tfrac{1}{n}\left(|\mathcal{P}_2|\right)$. From these results follows that the symmetric private capacity $P_{sym}\left(\mathcal{N}_{\mathcal{E}_2\mathcal{D}}\right)$ can be expressed as

$$P_{sym}\left(\mathcal{N}_{\mathcal{E}_2\mathcal{D}}\right) = C_{sym}\left(\mathcal{N}_{\mathcal{E}_1\mathcal{E}_2}\right) - C_{sym}\left(\mathcal{N}_{\mathcal{E}_1\mathcal{D}}\right), \quad (45)$$

thus for the channel between the quantum relay encoder $\mathcal{E}_2$ and Bob, the achievable symmetric private classical capacity is

$$P_{sym}\left(\mathcal{N}_{\mathcal{E}_2\mathcal{D}}\right) = \lim_{n\to\infty}\frac{1}{n}\left(\left|\mathcal{G}(\mathcal{N}_{phase},\beta)\right| - |\mathcal{P}_2|\right), \quad (46)$$

which is equal to

$$\begin{aligned}P_{sym}\left(\mathcal{N}_{\mathcal{E}_2\mathcal{D}}\right) &= \lim_{n\to\infty}\frac{1}{n}\left|\left(\mathcal{G}(\mathcal{N}_{phase},\beta)\right)\cap\left(\mathcal{G}(\mathcal{N}_{amp.},\beta)\right)\right| \\ &= \lim_{n\to\infty}\frac{1}{n}\left|S_{in}\right|.\end{aligned}$$
(47)

The private capacity which can be achieved over the quantum relay channel $\mathcal{N}_{\mathcal{E}_1\mathcal{E}_2\mathcal{D}}$ with the combination of the *superactivation-assistance* and *polar encoding* will be referred to as $P^*_{sym}\left(\mathcal{N}_{\mathcal{E}_1\mathcal{E}_2\mathcal{D}}\right)$.

## V. THEOREMS AND PROOFS ON THE QUANTUM RELAY CHANNEL

In this section we present and the theorems on the reliability and security of the superactivation-assisted quantum relay encoder.

**Theorem 3.** *The security of the quantum channels between Alice and the quantum relay encoder and the quantum relay encoder and Bob are guaranteed by the proposed quantum relay encoder scheme.*

*Proof.* Using our quantum polar codeword sets from (12), (14), (15) and (16) the *security* of the scheme is guaranteed, since the transmitted codewords on channels $\mathcal{N}_{\mathcal{E}_1\mathcal{E}_2}$ and $\mathcal{N}_{\mathcal{E}_2\mathcal{D}}$ are:

$$\begin{aligned}\mathcal{N}_{\mathcal{E}_1\mathcal{E}_2} &: \mathcal{G}(\mathcal{N}_{phase},\beta) = \mathcal{P}_2 \cup S_{in}, \\ \mathcal{N}_{\mathcal{E}_2\mathcal{D}} &: \mathcal{G}(\mathcal{N}_{amp},\beta) \cap \mathcal{G}(\mathcal{N}_{phase},\beta) = S_{in}.\end{aligned} \quad (48)$$

which means, that the outputs of $\mathcal{E}_1$ and $\mathcal{E}_2$ are those polar codewords which will be completely useless for Eve. As we show here, for the quantum relay channel $\mathcal{N}_{\mathcal{E}_1\mathcal{E}_2\mathcal{D}}$ with quantum channels $\mathcal{N}_{\mathcal{E}_1\mathcal{E}_2}$ and $\mathcal{N}_{\mathcal{E}_2\mathcal{D}}$,

$$C_{Eve}(\mathcal{N}_{\mathcal{E}_1\mathcal{E}_2\mathcal{D}}) = C_{Eve}(\mathcal{N}_{\mathcal{E}_1\mathcal{E}_2}) + C_{Eve}(\mathcal{N}_{\mathcal{E}_2\mathcal{D}}) = 0. \quad (49)$$

In the channel section $\mathcal{N}_{\mathcal{E}_1\mathcal{E}_2}$, the encoder $\mathcal{E}_1$ outputs a codeword from the set $\mathcal{P}_2 \cup S_{in}$. The quantum polar codewords $\mathcal{P}_2$ are 'partly good' for Eve, i.e., some information would be leaked to the eavesdropper from the set $\mathcal{P}_2 \cup S_{in}$, however from (19), (20) and (36) follows that our polar encoding scheme completely makes no possible to Eve to obtain any information. In the channel section $\mathcal{N}_{\mathcal{E}_2\mathcal{D}}$, the relay encoder $\mathcal{E}_2$ outputs a codeword from $S_{in}$, which will be completely useless for Eve, which trivially follows from (19) and (37). Combining these results with (36), we get the following result for Eve's channel for these polar sets. Since the $\mathcal{P}_2$ possible 'good' polar codewords for Eve are in the set $\mathcal{P}_1 \cup \mathcal{P}_2$, we can use the previously derived results for these sets from (36) and (20):

$$\begin{aligned}C_{Eve}(\mathcal{N}_{\mathcal{E}_1\mathcal{E}_2}) &= \lim_{n\to\infty}\frac{1}{n}|\mathcal{P}_2| + \lim_{n\to\infty}\frac{1}{n}|S_{in}| = 0, \\ C_{Eve}(\mathcal{N}_{\mathcal{E}_2\mathcal{D}}) &= \lim_{n\to\infty}\frac{1}{n}|S_{in}| = 0.\end{aligned} \quad (50)$$

As follows, from the set $\mathcal{P}_2$ the eavesdropper cannot obtain enough information in section $\mathcal{N}_{\mathcal{E}_1\mathcal{E}_2}$, nor in $\mathcal{N}_{\mathcal{E}_2\mathcal{D}}$, and using the proposed quantum polar codeword sets and our encoding scheme, and $C_{Eve}(\mathcal{N}_{\mathcal{E}_1\mathcal{E}_2\mathcal{D}}) = C_{Eve}(\mathcal{N}_{\mathcal{E}_1\mathcal{E}_2}) + C_{Eve}(\mathcal{N}_{\mathcal{E}_2\mathcal{D}}) = 0$. The condition $P_{sym}(\mathcal{N}_{\mathcal{E}_1\mathcal{E}_2\mathcal{D}}) > 0$ is also satisfied for the relay channel $\mathcal{N}_{\mathcal{E}_1\mathcal{E}_2\mathcal{D}}$.

∎

Our main result on the reliability of the superactivation-assisted quantum relay encoder is summarized in Theorem 4.

**Theorem 4.** *Using the unreliable quantum relay encoder $\mathcal{E}_2$ with $0 < p_{\varepsilon_2} < 0.5$, the superactivation-assisted private classical capacity $P_{sym}(\mathcal{N}_{\mathcal{E}_1\mathcal{E}_2\mathcal{D}})$ of the quantum relay channel $\mathcal{N}_{\mathcal{E}_1\mathcal{E}_2\mathcal{D}}$ will be positive and the reliability of the quantum relay encoder equals to $p_{\varepsilon_2} = 1$.*

*Proof:* First, Alice generates codeword $A$ with $\mathcal{E}_1$. In the next step, she transmits it over $\mathcal{N}_{\mathcal{E}_1\mathcal{E}_2}$ and feeds $B'$ into $\mathcal{E}_2$, which will result in $A'$. It will be transmitted over $\mathcal{N}_{\mathcal{E}_2\mathcal{D}}$, which will result in Bob's input $B$. For positive private classical capacity $P_{sym} > 0$, both the phase and the amplitude have to be transmitted; however, the encoders $\mathcal{E}_1$ and $\mathcal{E}_2$ individually cannot encode both of them. Using channel $\mathcal{N}_{\mathcal{E}_2\mathcal{D}}$ between $\mathcal{E}_2$ and $\mathcal{D}$, for the superactivation of we define the following channel $\mathcal{M}$

$$\mathcal{M} = p\mathcal{N}_{\mathcal{E}_2\mathcal{D}} \otimes |0\rangle\langle 0| + (1-p)\mathcal{A}_e \otimes |1\rangle\langle 1|, \quad (51)$$

where $0 \leq p \leq 1$ and $\mathcal{A}_e$ is the 50% erasure channel [5]. The channel $\mathcal{M}$ with probability $p$ is a $\mathcal{N}_{\mathcal{E}_2\mathcal{D}}$ channel (i.e., an $\mathcal{N}_{\mathcal{E}_1\mathcal{E}_2\mathcal{D}} = \mathcal{N}_{\mathcal{E}_1\mathcal{E}_2}\mathcal{N}_{\mathcal{E}_2\mathcal{D}}$ channel, since the encoder $\mathcal{E}_1$ is applied before channel $\mathcal{N}_{\mathcal{E}_2\mathcal{D}}$), otherwise, with probability $(1-p)$, it is an 50% erasure channel, which also has zero private capacity, i.e., $P(\mathcal{A}_e) = 0$. To superactivate the joint construction of $\mathcal{M}_1 \otimes \mathcal{M}_2$, Alice will feed the following entangled system to the inputs (denoted by $A$ and $C$) of the joint channel:

$$\begin{aligned}\rho_{AC} &= \frac{1}{2}\Big(|0\rangle\langle 0|_{A_1} \otimes |0\rangle\langle 0|_{C_1} + |0\rangle\langle 0|_{A_1} \otimes |0\rangle\langle 0|_{C_1}\Big), \\ &\otimes |\Psi_+\rangle\langle\Psi_+|\end{aligned} \quad (52)$$

where $|\Psi_+\rangle = \frac{1}{\sqrt{2}}(|00\rangle + |11\rangle)$. The $I_{coh}$ quantum coherent information of $\mathcal{M}_1 \otimes \mathcal{M}_2$ for the input system $\rho_{AC}$ is $I_{coh}(\mathcal{M}_1 \otimes \mathcal{M}_2) = 2p(1-p)I_{coh}(\mathcal{N}_{\mathcal{E}_1\mathcal{E}_2\mathcal{D}})$,

from which follows that for the private classical capacity of $\mathcal{M}_1 \otimes \mathcal{M}_2$:

$$P_{sym}(\mathcal{M}_1 \otimes \mathcal{M}_2) \geq 2p(1-p)P(\mathcal{N}_{\mathcal{E}_1\mathcal{E}_2\mathcal{D}}), \quad (53)$$

where $I_{coh}(\mathcal{N}_{\mathcal{E}_1\mathcal{E}_2\mathcal{D}})$ is the coherent information of channel $\mathcal{N}_{\mathcal{E}_1\mathcal{E}_2\mathcal{D}}$, and $0 < p < 1$. The lower bound on the achievable superactivated symmetric private classical capacity of $\mathcal{M}_1 \otimes \mathcal{M}_2$ is

$$P_{sym}(\mathcal{M}_1 \otimes \mathcal{M}_2) \geq \frac{1}{2} I_{coh}(\mathcal{N}_{\mathcal{E}_1\mathcal{E}_2\mathcal{D}}). \quad (54)$$

Using $P_{sym}(\mathcal{N}_{\mathcal{E}_1\mathcal{E}_2\mathcal{D}}) = I_{coh}(\mathcal{N}_{\mathcal{E}_1\mathcal{E}_2\mathcal{D}})$, we get the following lower bound for the $P$ symmetric private classical capacity of $\mathcal{M}_1 \otimes \mathcal{M}_2$:

$$P_{sym}(\mathcal{M}_1 \otimes \mathcal{M}_2) \geq \frac{1}{2} P_{sym}(\mathcal{N}_{\mathcal{E}_1\mathcal{E}_2\mathcal{D}}), \quad (55)$$

thus for our encoding scheme $P_{sym}(\mathcal{M}_1 \otimes \mathcal{M}_2) = \frac{1}{2} P_{sym}(\mathcal{N}_{\mathcal{E}_1\mathcal{E}_2\mathcal{D}})$. The required condition $I_{sym.}^{amp.}(A:B) > 0$ for the positive private capacity $P(\mathcal{N}_{\mathcal{E}_1\mathcal{E}_2\mathcal{D}})$ of the relay channel $\mathcal{N}_{\mathcal{E}_1\mathcal{E}_2\mathcal{D}}$ is also satisfied. Finally, the superactivation-assisted *single-use* private classical capacity $P_{sym}^{(1)}(\mathcal{M}_1 \otimes \mathcal{M}_2)$ of $\mathcal{M}_1 \otimes \mathcal{M}_2$ is evaluated as follows:

$$\begin{aligned} P_{sym}^{(1)}(\mathcal{M}_1 \otimes \mathcal{M}_2) &= \frac{1}{2}\Big(I_{sym.}^{phase}(A:B) - I(A:E)\Big) \\ &= \frac{1}{2}\begin{pmatrix} S\big((\sigma_0^{phase}+\sigma_1^{phase})/2\big) - S(\sigma_0^{phase}/2) \\ -S(\sigma_1^{phase}/2) - I(A:E) \end{pmatrix}, \end{aligned} \quad (56)$$

which is the half of the private classical capacity $P(\mathcal{N}_{\mathcal{E}_1\mathcal{E}_2\mathcal{D}})$, that can be achieved over $\mathcal{N}_{\mathcal{E}_1\mathcal{E}_2\mathcal{D}}$ if $p_{\varepsilon_2} = 1$. The following result concludes our proof, since $p_{\varepsilon_2} P_{sym}(\mathcal{N}_{\mathcal{E}_1\mathcal{E}_2\mathcal{D}}) < \frac{1}{2} P_{sym}(\mathcal{N}_{\mathcal{E}_1\mathcal{E}_2\mathcal{D}})$, if the $p_{\varepsilon_2}$ initial success probability of encoder $\mathcal{E}_2$ was between $0 < p_{\varepsilon_2} < 0.5$.

∎

In case of $0 < p_{\varepsilon_2} < 0.5$ then the superactivation-assistance of $\mathcal{E}_2$ can enhance the *reliability* the private classical communication over $\mathcal{N}_{\mathcal{E}_1\mathcal{E}_2\mathcal{D}}$ using the quantum relay encoder $\mathcal{E}_2$. Assuming asymptotic limit with $n \to \infty$, for the superactivation-assisted private classical capacity of $\mathcal{M}_1 \otimes \mathcal{M}_2$ the following relation holds:

$$P_{sym}(\mathcal{M}_1 \otimes \mathcal{M}_2) \geq P_{sym}^{(1)}(\mathcal{M}_1 \otimes \mathcal{M}_2). \quad (57)$$

The quantum relay encoder $\mathcal{E}_2$ with superactivation-assistance is illustrated in Fig. 2. The joint channel construction $\mathcal{M}_1 \otimes \mathcal{M}_2$ realizes the quantum relay encoder $\mathcal{E}_2$ with $p_{\varepsilon_2} = 1$. Using this scheme, the rate of private communication between Alice and Bob can be increased if initially the $p_{\varepsilon_2}$ success probability of $\mathcal{E}_2$ was $0 < p_{\varepsilon_2} < 0.5$, while the reliability of the quantum relay encoder can be maximized to the $p_{\varepsilon_2} = 1$. We use the same channel $\mathcal{M}$ as defined in (51), but in this case, instead of applying $\rho_{AC}$ in (52) Alice feeds to the inputs of $\mathcal{M}_1 \otimes \mathcal{M}_2$ an arbitrary quantum system $\rho \in \mathcal{G}(\mathcal{N}_{phase}, \beta)$ (assumed being symmetric in $A$ and $C$, which will result in $\mathcal{N}_{\mathcal{E}_1\mathcal{E}_2\mathcal{D}} \otimes \mathcal{A}_e = \mathcal{A}_e \otimes \mathcal{N}_{\mathcal{E}_1\mathcal{E}_2\mathcal{D}}$).

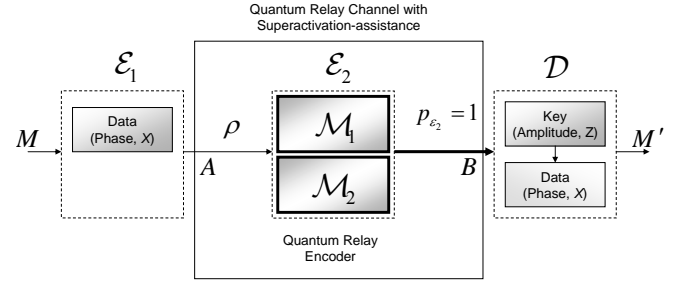

**Fig. 2.** The quantum relay encoder with superactivation-assistance. The rate of private communication can be increased if the initial reliability $p_{\varepsilon_2}$ was in $0 < p_{\varepsilon_2} < 0.5$. In the joint channel structure the reliability of the quantum relay encoder will be $p_{\varepsilon_2} = 1$.

Using our polar set construction the result of Theorem 5 is satisfied for the quantum relay encoder.

**Theorem 5.** *Using superactivation-assisted polar coding and a degraded quantum relay encoder $\mathcal{E}_2$ with $0 < p_{\varepsilon_2} < 0.5$ and input $\rho \in \mathcal{G}(\mathcal{N}_{phase}, \beta)$, for the superactivation-assisted private capacity $P_{sym}^{*}(\mathcal{N}_{\mathcal{E}_1\mathcal{E}_2\mathcal{D}})$ and the symmetric private classical capacity $P_{sym}(\mathcal{N}_{\mathcal{E}_1\mathcal{E}_2\mathcal{D}})$ hold that*

$$\frac{1}{2} P_{sym}^{*}(\mathcal{N}_{\mathcal{E}_1\mathcal{E}_2\mathcal{D}}) > \frac{1}{2} P_{sym}(\mathcal{N}_{\mathcal{E}_1\mathcal{E}_2\mathcal{D}}).$$

*Proof*: For the input system, the quantum coherent information of $\mathcal{M}_1 \otimes \mathcal{M}_2$ is evaluated as follows:

$$I_{coh}(\mathcal{M}_1 \otimes \mathcal{M}_2) = p^2 I_{coh}(\mathcal{N}_{\mathcal{E}_1 \mathcal{E}_2 \mathcal{D}} \otimes \mathcal{N}_{\mathcal{E}_1 \mathcal{E}_2 \mathcal{D}})$$
$$+ p(1-p) I_{coh}(\mathcal{N}_{\mathcal{E}_1 \mathcal{E}_2 \mathcal{D}}) + p(1-p) I_{coh}(\mathcal{N}_{\mathcal{E}_1 \mathcal{E}_2 \mathcal{D}}) \quad (58)$$
$$+ p(1-p) I_{coh}(\mathcal{A}_e \otimes \mathcal{A}_e),$$

where

$$I_{coh}(\mathcal{A}_e \otimes \mathcal{A}_e) = 0 \quad (59)$$

and

$$I_{coh}(\mathcal{N}_{\mathcal{E}_1 \mathcal{E}_2 \mathcal{D}} \otimes \mathcal{N}_{\mathcal{E}_1 \mathcal{E}_2 \mathcal{D}}) = 0, \quad (60)$$

since $\mathcal{A}_e \otimes \mathcal{A}_e$ is a symmetric channel [5]. Furthermore,

$$I_{coh}(\mathcal{N}_{\mathcal{E}_1 \mathcal{E}_2 \mathcal{D}} \otimes \mathcal{N}_{\mathcal{E}_1 \mathcal{E}_2 \mathcal{D}}) = 0, \quad (61)$$

since quantum relay encoder $\mathcal{E}_2$ can add the amplitude information to the phase information received from $\mathcal{E}_1$ in message $A$ only with probability $0 < p_{\mathcal{E}_2} < 0.5$. (The relay channel $\mathcal{N}_{\mathcal{E}_1 \mathcal{E}_2 \mathcal{D}}$ can transmit quantum information only with probability $p_{\mathcal{E}_2}$; otherwise it produces an output $\sigma$, which will result in zero quantum coherent information.) It trivially leads to zero quantum capacity $Q(\mathcal{N}_{\mathcal{E}_1 \mathcal{E}_2 \mathcal{D}} \otimes \mathcal{N}_{\mathcal{E}_1 \mathcal{E}_2 \mathcal{D}}) = 0$, since to achieve positive quantum capacity $p_{\mathcal{E}_2} > 0.5$ is required. ∎

In Theorem 6 we combine the results of Theorems 4 and 5.

**Theorem 6.** *Using the superactivation-assisted quantum relay channel $\mathcal{N}_{\mathcal{E}_1 \mathcal{E}_2 \mathcal{D}}$, the reliability of any $\mathcal{E}_2$ will be the maximal $p_{\mathcal{E}_2} = 1$ and the symmetric private classical capacity will be lower bounded by $P_{sym}^*(\mathcal{M}_1 \otimes \mathcal{M}_2) \geq \lim_{n \to \infty} \frac{1}{n}\left(\frac{1}{2}|S_{in}|\right)$.*

*Proof*: Assuming a quantum relay encoder $\mathcal{E}_2$ with reliability $p_{\mathcal{E}_2}$, this result reduces to $p_{\mathcal{E}_2} P_{sym}(\mathcal{N}_{\mathcal{E}_2 \mathcal{D}})$. Using the channel structure $\mathcal{M}_1 \otimes \mathcal{M}_2$ constructed for the superactivation of quantum relay encoder $\mathcal{E}_2$, using the result obtained in (58),

$$I_{coh}(\mathcal{M}_1 \otimes \mathcal{M}_2) = 2p(1-p) I_{coh}(\mathcal{N}_{\mathcal{E}_1 \mathcal{E}_2 \mathcal{D}}) \quad (62)$$

where $0 < p < 1$ and $I_{coh}(\mathcal{N}_{\mathcal{E}_1 \mathcal{E}_2 \mathcal{D}}) > 0$, combining with

$$P_{sym}^*(\mathcal{M}_1 \otimes \mathcal{M}_2) \geq \frac{1}{2} P_{sym}(\mathcal{N}_{\mathcal{E}_1 \mathcal{E}_2 \mathcal{D}}) \quad (63)$$

and using $P_{sym}(\mathcal{N}_{\mathcal{E}_1 \mathcal{E}_2 \mathcal{D}}) = I_{coh}(\mathcal{N}_{\mathcal{E}_1 \mathcal{E}_2 \mathcal{D}})$, lead us to the following result regarding the symmetric private classical capacity of superactivation-assisted polar encoding-based quantum relay channel $\mathcal{N}_{\mathcal{E}_1 \mathcal{E}_2 \mathcal{D}}$:

$P_{sym}^*(\mathcal{M}_1 \otimes \mathcal{M}_2) \geq \frac{1}{2} P_{sym}(\mathcal{N}_{\mathcal{E}_1 \mathcal{E}_2 \mathcal{D}})$. In the asymptotic limit with $n \to \infty$:

$$P_{sym}^*(\mathcal{M}_1 \otimes \mathcal{M}_2)^{\otimes n} \geq P_{sym}^*(\mathcal{M}_1 \otimes \mathcal{M}_2) \geq \lim_{n \to \infty} \frac{1}{n}\left(\frac{1}{2}|S_{in}|\right). \quad (64)$$

For the polar-coding based superactivation of relay encoder $\mathcal{E}_2$ our proof is concluded as follows:

$$P_{sym}^*(\mathcal{M}_1 \otimes \mathcal{M}_2)$$
$$\geq \lim_{n \to \infty} \frac{1}{n}\left(\frac{1}{2}\left|(\mathcal{G}(\mathcal{N}_{phase}, \beta)) \cap (\mathcal{G}(\mathcal{N}_{amp.}, \beta))\right|\right),$$
$$P_{sym}^*(\mathcal{M}_1 \otimes \mathcal{M}_2) \geq P_{sym}^*(\mathcal{M}_1 \otimes \mathcal{M}_2) \geq \lim_{n \to \infty} \frac{1}{n}\left(\frac{1}{2}|S_{in}|\right),$$

where $\lim_{n \to \infty} \frac{1}{n}|S_{in}| \geq P_{sym}(\mathcal{N}_{\mathcal{E}_1 \mathcal{E}_2 \mathcal{D}})$, since the maximum of the rate of any private communication over the relay channel $\mathcal{N}_{\mathcal{E}_1 \mathcal{E}_2 \mathcal{D}}$ cannot exceed $\lim_{n \to \infty} \frac{1}{n}|S_{in}|$, which concludes our proof. For the output $B^*$ of the superactivation-assisted quantum relay channel $\mathcal{N}_{\mathcal{E}_1 \mathcal{E}_2 \mathcal{D}}$:

$$|B^*| = \frac{1}{2}\left(|\mathcal{G}(\mathcal{N}_{phase}, \beta) \setminus \mathcal{P}_2|\right)$$
$$= \frac{1}{2}|S_{in}| > |B| = p_{\mathcal{E}_2}|S_{in}|, \quad (65)$$

i.e., if $0 < p_{\mathcal{E}_2} < 0.5$ the $P_{sym}(\mathcal{N}_{\mathcal{E}_1 \mathcal{E}_2 \mathcal{D}})$ private capacity of $\mathcal{N}_{\mathcal{E}_1 \mathcal{E}_2 \mathcal{D}}$ can be achieved by the combination of the proposed polar coding scheme and the superactivated quantum relay channel $\mathcal{N}_{\mathcal{E}_1 \mathcal{E}_2 \mathcal{D}}$. ∎

## VI. CONCLUSION

The polar coding is a revolutionary channel coding technique, which makes it possible to achieve the symmetric capacity of a noisy communication channel by the restructuring of the initial error probabilities. In the case of a quantum system, the problem is more complicated, since the error characteristic of a quantum communication channel significantly differs from the characteristic of quantum communication channels. In this paper we have shown that by combining the polar coding with superactivation-assistance, the reliability of the quantum relay encoder can be increased and the rate of the private communication over the superactivation-assisted relay quantum channel can be maximized at the same time. The proposed encoding scheme can be a useful tool in secure quantum communications.

ACKNOWLEDGMENT

The results discussed above are supported by the grant TAMOP-4.2.2.B-10/1--2010-0009 and COST Action MP1006.

# Appendix

## A. Polar Coding

Polar codes belong to the group of error-correcting codes [3]. The polar codewords can be used to achieve the symmetric capacity of classical discrete memoryless channels (DMCs). The basic idea behind the construction of polar codes is channel selection called polarization: assuming $n$ identical DMCs we can create two sets by means of an encoder. *Good* channels are nearly noiseless while *bad* channels have nearly zero capacity. Furthermore, for large enough $n$, the fraction of good channels approaches the symmetric capacity of the original DMC.

The polarization effect is represented by means of generator matrix $G_k$ having $k \times k$ of size [3] calculated in a recursive way

$$G_k = \left(I_{k/2} \otimes G_2\right) R_k \left(I_2 \otimes G_{k/2}\right), \quad (A.1)$$

where

$$G_2 = \begin{pmatrix} 1 & 1 \\ 0 & 1 \end{pmatrix} \quad (A.2)$$

and $I_{k/2}$ is the $\frac{k}{2} \times \frac{k}{2}$ identity matrix and $R_k$ is the $k \times k$ permutation operator. Now we present how matrix $G$ is related to the polarization effect. For an input message $M$ having $n = 2^k$ length, the encoded codeword $\rho_A$ is

$$\rho_A = f(M) = G_k M, \quad (A.3)$$

i.e., if $k = 2$, then

$$G \begin{pmatrix} M_1 \\ M_2 \end{pmatrix} = \begin{pmatrix} M_1 \oplus M_2 \\ M_2 \end{pmatrix}. \quad (A.4)$$

For the transmission of an *n*-length encoded codeword $A$, and *k*-level recursion with matrix $G_k$, using *n*-times the noisy quantum channel $\mathcal{N}$, the error-probability of the transmission is [3]

$$p_{error}(\rho_A) = n 2^{-n^\beta}. \quad (A.5)$$

In Fig. A.1(a) we show the first-level 'bad' channel $\mathcal{B}$, and in Fig. A.1(b) the first-level 'good' channel $\mathcal{G}$ is depicted [3], [6], [7]. The difference between the two channels is the knowledge of input $u_1$ on Bob's side. For the 'bad' channel $\mathcal{B}$ the input $u_1$ is unknown. In Fig. A.1(c), the second-level channel $\mathcal{N}_4$ is shown, which is the combination of the two first-level channels $\mathcal{N}_2$. From the two quantum channels, a new one, denoted by $\mathcal{N}_2$, is constructed by a simple CNOT gate. The recursion can be repeated over $k$ levels, with $n = 2^k$ channel uses. The two independent $\mathcal{N}_2$ channels are combined into a higher-level channel, denoted by $\mathcal{N}_4$. The scheme also contains a permutation operator $R$, which maps both the input and permutated inputs.

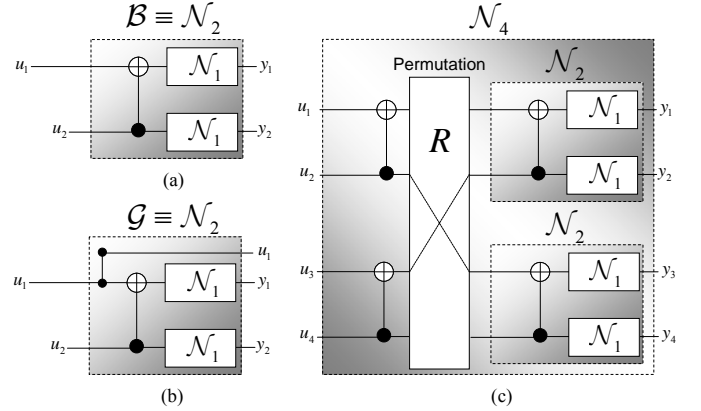

**Fig. A.1.** (a): The 'bad' channel: input $u_1$ is not known by Bob. (b): The 'good' channel: $u_1$ is also known on Bob's side. (c): The recursive channel construction from two lower-level channels. $R$ is the permutation operator.

The 'bad' and 'good' channels from Fig. A.1(a) and Fig. A.1(b) are defined as follows:

$$\mathcal{G} \equiv \mathcal{N}_2\left(u_1, y_1, y_2 \big| u_2\right) = \frac{1}{2} \mathcal{N}_1\left(y_1 \big| u_1 \oplus u_2\right) \mathcal{N}_1\left(y_2 \big| u_2\right),$$
$$\mathcal{B} \equiv \mathcal{N}_2\left(y_1, y_2 \big| u_1\right) = \frac{1}{2} \sum_{u_2 \in \{0,1\}} \mathcal{N}_1\left(y_1 \big| u_1 \oplus u_2\right) \mathcal{N}_1\left(y_2 \big| u_2\right).$$
$$(A.6)$$

## B. The Bhattacharya Parameter

In the polarization of channel $\mathcal{N}$, the Bhattacharya parameter $B(\mathcal{N})$ describes the noise of the transmission and it is analogous to the fidelity $F$ of the transmission [6], [7], and for output $y$ is defined as

$$B(\mathcal{N}) = \sum_{y} \sqrt{\mathcal{N}(y|0)\mathcal{N}(y|1)}. \quad (A.7)$$

If Alice chooses an *l*-length message *M* with *uniform* probability distribution from $\{0,1\}^l$ and encodes and sends it to Bob over the noisy quantum channel with *n* channel uses, then the probability that Bob's decoder will not fail

$$p(M' = M) \geq 1 - \sum_{i=1}^{n} B(\mathcal{N}_i), \quad (A.8)$$

Assuming that Alice's *l*-length message *M* is selected uniformly from a set of size *K* and transmitted by means of an *n*-length codeword, then the reliability of the transmission can be expressed by the average error probability $p(M' \neq M)$, as follows:

$$p(M' \neq M) = \frac{1}{K} \sum_{k=1}^{K} \Pr(M' \neq k | M = k), \quad (A.9)$$

where

$$\lim_{n \to \infty} p(M' \neq M) = 0. \quad (A.10)$$

Using a *k*-level structure, for a given set of channels indices $B(\mathcal{N}_i) = 0$ or $B(\mathcal{N}_i) = 1$, respectively. If the channel is "good," $\mathcal{G}(\mathcal{N}, \beta)$, then it can transmit perfectly the input, thus

$$B(\mathcal{N}_i) \to 0, \quad (A.11)$$

while if the channel belongs to the "bad" set, $\mathcal{B}(\mathcal{N}, \beta)$, then

$$B(\mathcal{N}_i) \to 1. \quad (A.12)$$

The two different values of $B(\mathcal{N}_i)$ indicate that for large enough uses *n* of the quantum channel structure $\mathcal{N}$, the channel structure $\mathcal{N}^{\otimes n}$ will be polarized. For the channels from set $\mathcal{G}(\mathcal{N}, \beta)$, $B(\mathcal{N}_i) \to 0$ while the capacity of the channels will be nearly 1. For the channels from set $\mathcal{B}(\mathcal{N}, \beta)$, $B(\mathcal{N}_i) \to 1$, i.e., the capacity of these channels will be nearly equal to 0.

### C. Channel Parameters

The symmetric classical capacity $C_{sym}(\mathcal{N}^{\otimes n})$ will be defined by the fraction of the good and bad channels, or in other words, the fraction of the Bhattacharya parameters of "good" and "bad" channels. In the polar encoding, for the $\mathcal{B}(\mathcal{N}, \beta)$ channels, Alice freezes the inputs, and valuable information will be transmitted only over the $\mathcal{G}(\mathcal{N}, \beta)$ channels. The freezing can be made by choosing a determined input to $\mathcal{B}(\mathcal{N}, \beta)$, for which the value is also known for Bob on the decoding side [3], [6], [7].

The following important result can be derived from (A.6) for the mutual information of these channels for uniform inputs $u_1$ and $u_2$, namely [3], [6], [7]:

$$\begin{aligned} &I(\mathcal{N}_2(y_1, y_2 | u_1)) + I(\mathcal{N}_2(u_1, y_1, y_2 | u_2)) \\ &= I(u_1 : y_1 y_2) + I(u_2 : u_1 y_1 y_2) \\ &= 2I(\mathcal{N}_1), \end{aligned} \quad (A.13)$$

where

$$I(u_1 : y_1 y_2) \leq I(\mathcal{N}_1) \quad (A.14)$$

and

$$I(u_2 : u_1 y_1 y_2) \geq I(\mathcal{N}_1), \quad (A.15)$$

i.e.,

$$\begin{aligned} I(u_1 : y_1 y_2) &\leq I(\mathcal{N}_1) \leq I(u_2 : u_1 y_1 y_2) \\ I(\mathcal{N}_2(y_1, y_2 | u_1)) &\leq I(\mathcal{N}_1) \leq I(\mathcal{N}_2(u_1, y_1, y_2 | u_2)). \end{aligned} \quad (A.16)$$

For the polarized 'bad' $\mathcal{B}(\mathcal{N}, \beta)$ and 'good' $\mathcal{G}(\mathcal{N}, \beta)$ channels the following rules hold:

$$\begin{aligned} \mathcal{G}(\mathcal{N}, \beta) &\equiv \left\{ i \in n : B(\mathcal{N}_i^{\otimes n}) < \frac{1}{n} 2^{-n^\beta} \right\}, \\ \mathcal{B}(\mathcal{N}, \beta) &= [n] \setminus \mathcal{G}(\mathcal{N}_i, \beta). \end{aligned} \quad (A.17)$$

In (A.17), parameter $\beta$ is defined as

$$\beta = \frac{1}{n} \sum_{i=1}^{n} \log_n d_i, \quad (A.18)$$

where $d_i = d_{\min}(\mathbf{g_i}, \mathbf{g_{i+1}}, \ldots, \mathbf{g_n})$, and $g_i$ is the $i^{\text{th}}$ row vector of matrix $G_n$ As was shown [3], $\beta \leq 0.5$ if $k < 15$, while for $n \geq 16$: $\beta > 0.5$, along with

$$\lim_{n \to \infty} \frac{1}{n} \beta = 1. \quad (A.19)$$

Private classical communication over these structures means the following: in her message *A*, Alice sends her encoded private message $M$ only over channels $\mathcal{G}(\mathcal{N},\beta)$, while the remaining parts of $A$ are transmitted via $\mathcal{B}(\mathcal{N},\beta)$. Moreover, after the channels are being polarized, the fraction of $\mathcal{G}(\mathcal{N},\beta)$ and $\mathcal{B}(\mathcal{N},\beta)$ will be equal to the symmetric private classical capacity $P_{sym}(\mathcal{N})$. In our case, the input quantum channels $\mathcal{N}^{\otimes n}$ are insecure, i.e., they cannot transmit the amplitude and phase information simultaneously; however, using polar encoding, the parties will be able send both the amplitude and the phase over $\mathcal{N}^{\otimes n}$.

For the private communication scenario (see Fig. 1) the polarized channel construction, assuming a sufficient number of channel-use $n$ and $\beta < 0.5$, the following relation holds for a given set of codewords:

$$S_{Bob}(\mathcal{N}_{Bob},\beta) \equiv \left\{ i \in n : B(\mathcal{N}_{i,Bob}) < \frac{1}{n} 2^{-n^\beta} \right\},$$
$$S_{Eve}(\mathcal{N}_{Eve},\beta) \equiv \left\{ i \in n : B(\mathcal{N}_{i,Eve}) \geq 1 - \frac{1}{n} 2^{-n^\beta} \right\}.$$
(A.20)

For these codewords, the $R_{sym}$ achievable symmetric private classical communication rate assuming non-symmetric channel $\mathcal{N}_{Eve}$ (i.e., in this case the *maximization is needed*) can be expressed as the difference of the two mutual information functions,

$$R_{sym} = \lim_{n \to \infty} \frac{1}{n} \max_{all \ p_i, \rho_i} I(A:B) - I(A:E), \quad \text{(A.21)}$$

where the channel between Alice and Bob is symmetric. Using encoding scheme, the rate of the symmetric private classical capacity is

$$R_{sym}(\mathcal{N}) = \lim_{n \to \infty} \frac{1}{n} \max_{all \ p_i, \rho_i} \left( C(\mathcal{N}_{Bob}) - C(\mathcal{N}_{Eve}) \right)$$
$$= \lim_{n \to \infty} \frac{1}{n} \max_{all \ p_i, \rho_i} \left( I^{phase}_{sym.}(A:B) - I(A:E) \right).$$
(A.22)

The convergence of the Bhattacharya parameters of the $\mathcal{G}(\mathcal{N},\beta)$ "good" and $\mathcal{B}(\mathcal{N},\beta)$ "bad" channel sets, and the steps of the iteration process was proven by Arikan [3]. In the description of the convergence of the iteration, another important parameter was also introduced, namely the $L$ likelihood ratio, which is used for the description of Bob's decoding strategy.

### D. The Likelihood Parameter

The $L$ likelihood parameter measures whether the original input on Alice side was 0 or 1. From this value, Bob can decide with high probability, whether Alice sent 0 or 1. The likelihood parameter-based iteration is optimal and can be achieved in $\mathcal{O}(n \log n)$ time. For the quantum channel $\mathcal{N}_1$ with a given $n$-length output codeword $y$, the likelihood ratio [3], [6], [7] of the channel is defined as

$$L(\mathcal{N}_1, y) = \frac{\mathcal{N}_1(y|0)}{\mathcal{N}_1(y|1)}. \quad \text{(A.23)}$$

The likelihood parameter is computed for each of the $n$ bits of the received codeword $y$. The $L(\mathcal{N}_1, y)$ can be defined in a different way for the sets $\mathcal{G}(\mathcal{N},\beta)$ and $\mathcal{B}(\mathcal{N},\beta)$, since for the "bad" channels (see Fig. A.1(b)), it can be expressed as

$$L(\mathcal{B}(\mathcal{N},\beta), y_1, y_2) = \frac{1 + L(\mathcal{N}_1, y_1) L(\mathcal{N}_1, y_2)}{L(\mathcal{N}_1, y_1) + L(\mathcal{N}_1, y_2)}$$
$$= \frac{1 + \left( \frac{\mathcal{N}_1(y_1|0)}{\mathcal{N}_1(y_1|1)} \right) \left( \frac{\mathcal{N}_1(y_2|0)}{\mathcal{N}_1(y_2|1)} \right)}{\frac{\mathcal{N}_1(y_1|0)}{\mathcal{N}_1(y_1|1)} + \frac{\mathcal{N}_1(y_2|0)}{\mathcal{N}_1(y_2|1)}}. \quad \text{(A.24)}$$

For the set $\mathcal{G}(\mathcal{N},\beta)$, we have to distinguish the case input $u_1 = 0$ and $u_1 = 1$, since for the "good" channels, $u_1$ is also known at Bob's side (see Fig. A.1(b)). For $\mathcal{B}(\mathcal{N},\beta)$ with $u_1 = 0$,

$$L(\mathcal{B}(\mathcal{N},\beta), y_1, y_2) = L(\mathcal{N}_1, y_1) L(\mathcal{N}_1, y_2)$$
$$= \frac{\mathcal{N}_1(y_1|0)\mathcal{N}_1(y_2|0)}{\mathcal{N}_1(y_1|1)\mathcal{N}_1(y_2|1)}, \quad \text{(A.25)}$$

while for $u_1 = 1$,

$$L(\mathcal{B}(\mathcal{N},\beta), y_1, y_2) = \frac{L(\mathcal{N}_1, y_2)}{L(\mathcal{N}_1, y_1)}$$
$$= \frac{\mathcal{N}_1(y_2|0)\mathcal{N}_1(y_1|1)}{\mathcal{N}_1(y_2|1)\mathcal{N}_1(y_1|0)}. \quad \text{(A.26)}$$

Depending on the value of $L$, Bob will decide as follows. If $L > 0$, then he will decide for 0, since in this case 0 is more likely than 1. If $L < 0$, Bob will decide for 1. In the process of likelihood computing, Bob does not have to calculate the likelihood for the frozen input bits, since these bits are already known on his side.